\documentclass[aps,preprint]{revtex4}
\usepackage{epsfig,amsmath}
\usepackage{subfigure}
\usepackage{graphicx}
\usepackage{dcolumn}
\usepackage{stmaryrd}
\usepackage{mathrsfs}
\usepackage{pifont}
\usepackage{amsthm}
\usepackage{amssymb}
\usepackage{bm}
\usepackage{latexsym}
\usepackage[colorlinks=true,linkcolor=blue,citecolor=blue]{hyperref}
\usepackage{color}

\begin{document}

\title{Quantum algorithm for preparing the ground state of a system via
resonance transition}
\author{Hefeng Wang$^{1,2}$}
\thanks{email: wanghf@mail.xjtu.edu.cn}
\affiliation{$^{1}$Department of Applied Physics, Xi'an Jiaotong University, Xi'an 710049, China\\
$^{2}$Key Laboratory of Quantum Information and Quantum Optoelectronic
Devices, Shaanxi Province, China\\}
\date{\today }

\begin{abstract}
Preparing the ground state of a system is an important task in physics. We propose a quantum algorithm for preparing the ground state of a physical system that can be simulated on a quantum computer. The system is
coupled to an ancillary qubit, by introducing a resonance mechanism between
the ancilla qubit and the system, and combined with measurements performed
on the ancilla qubit, the system can be evolved to monotonically converge to
its ground state through an iterative procedure. We have simulated the
application of this algorithm for the Afflect-Kennedy-Lieb-Tasaki model,
whose ground state can be used as resource state in one-way quantum
computation.
\end{abstract}

\maketitle

Purification of quantum states is the key for many quantum applications,
e.g., highly-purified quantum states are required in improving the signal to
noise ratio in spectroscopy~\cite{snr1,snr2} and the resolution in metrology and quantum sensing~\cite{met1, met2, met3, met4}. It is also essential in
quantum information science, such as initializing a set of qubits to a known
state in many quantum algorithms, preparing resource state in one-way
quantum computation~\cite{raussendorf}, and supplying fresh ancillary qubits
in fault-tolerant quantum computing and quantum error correction~\cite{ft1,
ft2}. To purify a quantum state, in addition to physical cooling,
algorithmic cooling can be used to reduce the entropy of the system. In
quantum computation, algorithmic cooling can be used for preparing the
ground state of a quantum computer by means of the computer itself~\cite%
{baugh, brassard}.

There are a few quantum cooling algorithms have been proposed. In Ref.~\cite%
{HBAC}, a heat-bath algorithmic cooling~(HBAC) approach was proposed and
demonstrated experimentally. In this approach, the entropy of qubits is
reduced by distributing more entropy to one of the qubits, which can release
the excess entropy to a heat bath through thermalization. HBAC is not a
universal cooling algorithm, it is mainly used for preparing polarized spins
as initial states for quantum computation. Another approach~\cite{DOS1, DOS2}
for quantum cooling is to engineer dissipative open-system dynamics to drive
quantum states to the ground state of a simulated system. This approach is
based on simulation of the Lindblad master equations, and is restricted to
the frustration-free Hamiltonians. In Ref.~\cite{DLAC}, a universal quantum
cooling approach was proposed, this approach can be applied to enhance the
probability of the ground-state projection through non-unitary operations
introduced through measurements. In this approach, it cannot be verified
directly that the system is cooled to its ground state. In Ref.~\cite%
{nakazato}, a method is proposed for purifying a quantum system \textit{A} to a pure
state through Zeno-like measurements on another quantum system \textit{B}
that is coupled to \textit{A}. The effect of performing a
series of frequent measurements on system \textit{B} introduces non-unitary
operations on system \textit{A} and drives it to a pure state.

In this work, we propose a quantum algorithm for preparing the ground state
of a system via resonance transition, provided that the ground state energy of the system is known. The system is coupled to an ancillary qubit and a resonance
transition is introduced between them, through an energy exchange with the
ancilla qubit, the system can be driven to monotonically converge to its
ground state in an iterative way. In this algorithm, the system is prepared
in an initial state and the amplitude of the ground state is amplified
through a resonance mechanism, by performing measurements on the ancilla
qubit, the system is purified to converge quickly to its ground state. The
algorithm can be applied to any system with a Hamiltonian that can be
simulated on a quantum computer. We have simulated the application of this
algorithm for the Afflect-Kennedy-Lieb-Tasaki~(AKLT) model, whose ground state can be used as resource state in one-way quantum computation.

\noindent \textbf{{\large {Results}}} \newline
\noindent \textbf{The Algorithm.} For a qubit coupled to a physical system,
when the qubit resonates with a transition in the system, it exhibits a
dynamical response and an energy exchange occurs between the qubit and the
system. By performing measurements on the qubit, the system can be purified to its ground state. Based on this, we propose a quantum algorithm for preparing the ground state of a system, provided that the ground state energy is known. For some systems, such as some oracle-based problems in quantum computation, the ground state energies are already known, or they can be obtained through the algorithm we introduced here and in ref.~\cite{wan}. In this algorithm, the system is coupled with an ancillary qubit and is prepared in an initial state which can be spanned by the eigen-basis of the Hamiltonian of the system, the amplitude of the ground state is amplified through a resonance mechanism. By performing measurements on the ancilla qubit, which introduces a non-unitary operations on the system~\cite{nakazato}, the system can be purified and driven to monotonically converge to its ground state. Details of the algorithm are as follows.

The algorithm requires $\left(n+2\right) $ qubits, two ancillary qubits and
$n$ qubits that represent the system. The first ancilla qubit is coupled to
a quantum register $R$ of $\left( n+1\right) $ qubits consisting of the
second ancilla qubit and an $n$-qubit quantum register representing a system
of dimension $N=2^{n}$. The Hamiltonian of the algorithm is constructed as%
\begin{equation}
H=-\frac{1}{2}\sigma _{z}\otimes I_{2}^{\otimes (n+1)}+I_{2}\otimes
H_{R}+c\sigma _{x}\otimes \sigma _{x}\otimes I_{N},
\end{equation}%
where
\begin{equation}
H_{R}=\varepsilon _{0}|0\rangle \langle 0|\otimes I_{N}+|1\rangle \langle
1|\otimes H_{S},
\end{equation}%
and $H_{S}$ is the Hamiltonian of the physical system. The eigenstates of
the system are $|\chi _{j}\rangle $ and $H_{S}|\chi _{j}\rangle =E_{j}|\chi
_{j}\rangle $ ($j=1$, $2$, $\ldots $, $N$), where $E_{j}$ are the
eigenvalues of $H_{S}$. $I_{2}$ is the two-dimensional identity operator, $%
\sigma _{x}$ and $\sigma _{z}$ are the Pauli matrices. The first term in
Eq.~($1$) is the Hamiltonian of the first ancilla qubit, the second term is
the Hamiltonian of the register $R$, and the third term describes the
interaction between the ancilla qubit and $R$. Here, $\varepsilon _{0}$ is a
reference parameter, and $c$ is the coupling strength between the first
ancilla qubit and $R$.

In the algorithm, a guess state $|\varphi ^{(0)}\rangle $ of the ground state of the system is prepared as the initial input state. The register $R$
of the circuit is prepared in state $|0\rangle |\varphi ^{(0)}\rangle $,
which is an eigenstate of the Hamiltonian $H_{R}$ with eigenvalue $%
\varepsilon _{0}$. We set the parameter $\varepsilon _{0}$ such that $%
\varepsilon _{0}-E_{1}=1$, where $E_{1}$ represents the ground state energy
of the system. The procedures of the algorithm are as follows:

For $k=1$:

$\left( i\right) $ Prepare the first ancilla qubit in its ground state $%
|0\rangle $ and the register $R$ in state $|0\rangle |\varphi
^{(k-1)}\rangle $.

$\left( ii\right) $ Construct Hamiltonian of the algorithm as shown in Eq.~($1$), and implement the time evolution operator $U(\tau )=\exp \left( -iH\tau\right) $ with
$\tau =\pi /(2c)$.

$\left( iii\right) $ Perform a measurement on the first ancilla qubit in its
computational basis. If the measurement result is in its excited state $|1\rangle $, go to the next step; otherwise, set $k=1$ and run the algorithm start from step $(i)$ again.

$\left( iv\right) $ Take the state of the last $n$ qubits obtained from step $%
(iii)$ as input state $|\varphi ^{(k)}\rangle $ for the system. Set $k=k+1$ and run the procedures from step $(i)$ again.

Here, $k$ represents the iteration number of the algorithm. The state $|\varphi ^{(m)}\rangle $ obtained on the last $n$ qubits is close to the ground state of the system. The time evolution operator $U(\tau )=\exp\left(-iH\tau \right)$ can be implemented efficiently through the Trotter formula~\cite{nc}.

The initial state of the system $|\varphi ^{(0)}\rangle $ can be spanned by the complete set of eigenstates of the system Hamiltonian $\{|\chi _{j}\rangle $, $j=1$, $2$, $\cdots $, $N\}$ as $|\varphi ^{(0)}\rangle =\sum\nolimits_{j=1}^{N}d_{j}|\chi _{j}\rangle $, where $d_{j}=\langle
\varphi ^{(0)}|\chi _{j}\rangle $ and $\sum\nolimits_{j=1}^{N}|d_{j}|^{2}=1$%
. In basis \{$|\Psi _{j}\rangle =|0\rangle |0\rangle |\chi _{j}\rangle $, $%
|\Psi _{N+j}\rangle =|1\rangle |1\rangle |\chi _{j}\rangle $, $j=1$, $2$, $%
\ldots $, $N$\}, the Hamiltonian of the algorithm can be decomposed as
direct sum of $N$ two-dimensional matrix as $H=\oplus _{j=1}^{N}H_{j}$. With
$\varepsilon _{0}-E_{1}=\omega =1$, $H_{j}$ are in the form
\begin{equation}
H_{j}=\left(
\begin{array}{cc}
\frac{1}{2}+E_{1} & c \\
c & \frac{1}{2}+E_{j}%
\end{array}%
\right).
\end{equation}

In the first iteration of the algorithm, the initial state of the circuit is $|\Psi _{0}\rangle =|0\rangle |0\rangle |\varphi ^{(0)}\rangle $, the unitary
evolution of the circuit is $U|\Psi _{0}\rangle
=\sum\nolimits_{j=1}^{N}d_{j}U_{j}|0\rangle |0\rangle |\chi _{j}\rangle $,
where $U_{j}=\exp (-iH_{j}\tau )$. Then we have
\begin{equation}
U_{1}|0\rangle |0\rangle |\chi _{1}\rangle =c_{1}|1\rangle |1\rangle |\chi
_{1}\rangle,
\end{equation}%
where $c_{1}=e^{-i(\alpha +\frac{\pi }{2})}$\ and $\alpha =\frac{2E_{1}+1}{4c%
}\pi $. And
\begin{equation}
U_{j}|0\rangle |0\rangle |\chi _{j}\rangle =c_{j0}|0\rangle |0\rangle |\chi
_{j}\rangle +c_{j1}|1\rangle |1\rangle |\chi _{j}\rangle ,\text{ \ \ }j>1
\end{equation}%
where
\begin{equation}
c_{j0}=\left[ \frac{1}{2}-\frac{\delta _{j}}{2\sqrt{4c^{2}+\delta _{j}{}^{2}}%
}+\left( \frac{1}{2}+\frac{\delta _{j}}{2\sqrt{4c^{2}+\delta _{j}{}^{2}}}%
\right) e^{\frac{i\pi \sqrt{4c^{2}+\delta _{j}{}^{2}}}{2c}}\right] \times
e^{-i\pi \frac{\kappa _{j}+\sqrt{4c^{2}+\delta _{j}{}^{2}}}{4c}},
\end{equation}%
\begin{equation}
c_{j1}=\frac{c}{\sqrt{4c^{2}+\delta _{j}{}^{2}}}e^{-i\pi \frac{\kappa _{j}+%
\sqrt{4c^{2}+\delta _{j}{}^{2}}}{4c}}\left( 1-e^{\frac{i\pi \sqrt{%
4c^{2}+\delta _{j}{}^{2}}}{2c}}\right)
\end{equation}%
and $\delta _{j}=E_{j}-E_{1}$, $\kappa _{j}=E_{1}+E_{j}+1$. Then
\begin{eqnarray}
U|\Psi _{0}\rangle &=&U|0\rangle |0\rangle \sum\nolimits_{j=1}^{N}d_{j}|\chi
_{j}\rangle =\sum\nolimits_{j=1}^{N}d_{j}U_{j}|0\rangle |0\rangle |\chi
_{j}\rangle  \notag \\
&=&|1\rangle \left( d_{1}c_{1}|1\rangle |\chi _{1}\rangle
+\sum\nolimits_{j=2}^{N}d_{j}c_{j1}|1\rangle |\chi _{j}\rangle \right)
\notag \\
&&+|0\rangle \sum\nolimits_{j=2}^{N}d_{j}c_{j0}|0\rangle |\chi _{j}\rangle .
\end{eqnarray}%
From Eq.~($7$), we can see that $\left\vert c_{j1}\right\vert ^{2}=\frac{%
c^{2}}{4c^{2}+\delta _{j}^{2}}\left[ 2-2\cos \left( \frac{\pi \sqrt{%
4c^{2}+\delta _{j}{}^{2}}}{2c}\right) \right] $. The probability of
the measurement on the ancilla qubit being in its excited state $|1\rangle $, depends on the coupling coefficient $c$ and the energy gaps between the ground state and the excited states $\delta _{j}$. In the case where $\delta _{j}\gg c$ and $c\ll 1$, $\left\vert c_{j1}\right\vert \ll 1$ and $\left\vert c_{j0}\right\vert \approx 1$. Therefore in the first iteration of the algorithm, the probability of the measurement on the ancilla qubit being in its excited state is $\approx |d_{1}|^{2}$. The system is evolved to the excited states of the system energy when the measurement on the ancilla qubit is in state $|0\rangle $, while the system is purified to a state that is close to its ground state when the measurement is in its excited state $|1\rangle $. When the measurement on the first ancilla qubit is in state $|1\rangle $, the circuit is collapsed to state $|\Psi ^{(1)}\rangle =\frac{1}{\sqrt{N_{r}}}|1\rangle \left(
d_{1}c_{1}|1\rangle |\chi _{1}\rangle
+\sum\nolimits_{j=2}^{N}d_{j}c_{j1}|1\rangle |\chi _{j}\rangle \right) $,
where $N_{r}=|d_{1}c_{1}|^{2}+\sum\nolimits_{j=2}^{N}|d_{j}c_{j1}|^{2}$ is
the renormalization factor. Because $|c_{1}|=1$ and $\left\vert
c_{j1}\right\vert \approx c\ll 1$ provided that $\delta _{j}\gg c$, the
state $|1\rangle |1\rangle |\chi _{1}\rangle $ that encodes the ground state
of $H_{S}$ contributes most to the state $|\Psi ^{(1)}\rangle $, as long as $
d_{1}$ is polynomial large. Ignoring phase factors, the state on the last $n$
qubits of the circuit can be written as $\frac{1}{\sqrt{1+(a_{0}c)^{2}}}%
\left( |\chi _{1}\rangle +a_{0}c|\bar{\chi _{1}}\rangle \right) $, where $%
a_{0}=\sqrt{\sum\nolimits_{j=2}^{N}\left\vert \frac{d_{j}c_{j1}}{d_{1}c}%
\right\vert ^{2}}$, $|\bar{\chi _{1}}\rangle =\frac{1}{a_{0}c}%
\sum\nolimits_{j=2}^{N}\frac{d_{j}}{\left\vert d_{1}\right\vert }%
c_{j1}|1\rangle |\chi _{j}\rangle $ represents a state that does not contain
the ground state $|\chi _{1}\rangle $. The amplitude of the ground state of the system is amplified in a rate about $1:a_{0}c$ through the resonance mechanism. After $m$ continuous measurements on the
first ancilla qubit being in its excited state $|1\rangle $, the system is
purified to state $\frac{1}{\sqrt{1+(a_{0}c)^{2m}}}\left[ |\chi _{1}\rangle
+\left( a_{0}c\right) ^{m}|\bar{\chi _{1}}\rangle \right] $, and the
amplitude of the component that does not contain the ground state of $H_{S}$
is compressed to be exponentially small with $m$.

The success probability for $\left( m+1\right) $ continuous measurements on
the first ancilla qubit to be in its excited state $|1\rangle $ is
\begin{eqnarray}
P_{\text{succ}} &=&|d_{1}|^{2}\prod_{k=1}^{m}\frac{1}{1+(a_{0}c)^{2k}}
\notag \\
&>&|d_{1}|^{2}\left[ \frac{1}{1+(a_{0}c)^{2}}\right] ^{m}  \notag \\
&\approx &|d_{1}|^{2}\left[ 1-(a_{0}c)^{2}\right] ^{m}.
\end{eqnarray}%
If the coupling coefficient $c$ is set such that $a_{0}c<1/\sqrt{m}$ and $%
\delta _{j}\gg c$, then the success probability of the algorithm $P_{\text{%
succ}}>|d_{1}|^{2}/e$ in the asymptotic limit of $m$. The system is purified
to a state that has fidelity of $(1-m\left( a_{0}c\right)^{2m})$ with the
ground state of the Hamiltonian $H_{S}$. The number of trials the algorithm has to be run is proportional to $1/P_{\text{succ}}$. The evolution time of the algorithm is $\tau=\pi /(2c)$, if the energy gap between the ground state and the excited states $\delta _{j}$ is polynomially large, then the coupling coefficient $c$ can be set to be polynomially small and $\delta _{j}\gg c$, then as long as $ |d_{1}|^{2}$ is not exponentially small, the algorithm can be run in polynomial time with finite success probability.

Since the amplitude of the ground state of the system is amplified in a rate about $1:a_{0}c$ where $a_{0}c \ll 1$, the system is evolved very fast to a state that is very close to the ground state of the system when the measurement on the first ancilla qubit is in its excited state. The success probability of the algorithm mainly depends on the overlap between the initial state and the ground state of the system. Once the measurement on the ancilla qubit fails to be in its excited state, we can start over the algorithm again. Then after a few iterations, the initial state is purified to be close to the ground state of the system. In practice, the system can be purified to a state that is very close to the ground state of the system in a few iterations of the algorithm, as we can see in the example below.

We have to implement the time evolution operator $U(\tau )=\exp \left(
-iH\tau \right) $. In the algorithm Hamiltonian $H$, as shown in Eq.~($1$),
the first two terms commute, while they do not commute with the third term.
The operator $U(\tau )$ can be implemented through the Trotter formula~\cite%
{nc}:
\begin{equation}
U(\tau )\!=\!\left[ e^{-i\left( -\frac{1}{2}\sigma _{z}\otimes
I_{2}^{\otimes (n+1)}+I_{2}\otimes H_{R}\right) \tau /L}e^{-i\left( c\sigma
_{x}\otimes \sigma _{x}\otimes I_{N}\right) \tau /L}\right]
^{L}\!+\!O\!\left( 1/L\right)
\end{equation}%
By making $L$ very large, the error can be made as small as possible. $L$
can be made sufficiently large such that the error is bounded by some
threshold. By applying the Trotter approximation, the evolution operator on the circuit is $U(\tau )-O\!\left( 1/L\right) $, which introduces a slight deviation from the real unitary evolution of the state in an iteration of the algorithm. Then the probability for $\left( m+1\right) $ continuous measurements on the first ancilla qubit to be in its excited state $|1\rangle $ becomes $P_{\text{succ}}^{\prime
}\approx |d_{1}|^{2}\left[ 1-(a_{0}c)^{2}\right] ^{m}\left[ 1-O\!\left(
1/L\right) \right] ^{m+1}$.

A slight modification of the algorithm can be used for obtaining the ground
state energy of a system. When the transition frequency between the
reference state and an eigenstate of the system matches the frequency of the
first ancilla qubit, it contributes the most to the excitation of the qubit. By
performing measurements on the first ancilla qubit to obtain its excitation probability, a peak in the excitation rate of the qubit will be observed. Therefore by varying the eigenvalue of the reference state $\varepsilon _{0}$, and run the algorithm, we can locate the transition frequency between the reference state and the ground state and
obtain the ground state energy of the system. The procedures are the same as in Ref.~\cite{wan}.

\noindent \textbf{Simulating the algorithm for the AKLT model.} In the
following, we simulate the algorithm for the one-dimensional
Afflect-Kennedy-Lieb-Tasaki model. The ground state of the
two-dimensional AKLT model can be used as resource state for universal
one-way quantum computation~\cite{AKLT}. The AKLT model is a gapped model,
therefore it can be cooled to its ground state, and by performing
single-qubit operations, one-way quantum computation can be implemented on
this model. In Ref.~\cite{lavoie}, the authors simulated one-way quantum
computation on the one-dimensional AKLT model by preparing the solid bond
state of the model on photon system, but they did not use a cooling method~%
\cite{raussendorf}. Here, by applying the algorithm we proposed, we can evolve the simulated AKLT model on qubit system to its ground state.

The one-dimensional AKLT model consists of a linear chain of $N$ spin-$1$'s
in the bulk and two spin-$1/2$'s on the boundary. The spin-$1$ operators are
represented by $\vec{S}_{k}$\ and spin-$1/2$ operators by $\vec{s}_{j}$,
where $j=0,N+1$. The Hamiltonian of the system is~\cite{fan}%
\begin{equation}
H_{\text{AKLT}}=\sum_{k=1}^{N-1}\left[ \vec{S}_{k}\vec{S}_{k+1}+\frac{1}{3}%
\left( \vec{S}_{k}\vec{S}_{k+1}\right) ^{2}\right] +\theta _{0,1}+\theta
_{N,N+1}.
\end{equation}%
The boundary terms $\theta $ describe interaction between a spin-$1/2$ and a
spin-$1$, and
\begin{equation}
\theta _{0,1}=\frac{2}{3}\left( 1+\vec{s}_{0}\vec{S}_{1}\right),\text{\ \ }%
\theta _{N,N+1}=\frac{2}{3}\left( 1+\vec{s}_{N+1}\vec{S}_{N}\right).
\end{equation}%
Using the $3$-spin $1d$-AKLT model as an example, we simulate the algorithm for
preparing the ground state of this system. The Hamiltonian of the system is%
\begin{equation}
H_{S}=\frac{2}{3}\left( 1+\vec{s}_{0}\vec{S}_{1}\right) +\frac{2}{3}\left( 1+%
\vec{s}_{2}\vec{S}_{1}\right).
\end{equation}%
The ground state of the system Hamiltonian $H_{S}$ is unique and the
eigenvalue is zero. We use four qubits to simulate this system. The states $%
|-1\rangle $, $|0\rangle $, $|1\rangle $ of the spin-$1$ system are
represented in the symmetric space of two qubits:
\begin{equation}
|1,1\rangle =|00\rangle, |1,0\rangle =\frac{1}{\sqrt{2}}\left( |01\rangle
+|10\rangle \right), |1,-1\rangle = |11\rangle .
\end{equation}

We set the eigenvalue of the reference state as $\varepsilon _{0}=1$. In the basis above, the ground state of the simulated AKLT model in qubit system is $1/\sqrt{12}\left(
|0011\rangle +|1100\rangle +|0101\rangle +|1010\rangle \right) -1/\sqrt{3}%
\left( |0110\rangle +|1001\rangle \right) $. The initial state of the system is prepared in state $|1100\rangle $, which has fidelity of $1/12$ with the ground state of the system.

In the following, we show that the ground state energy of the AKLT model above can be obtained using the algorithm we proposed here. The coupling coefficient is set as $c=0.05$. We vary the eigenvalue of the reference state $\varepsilon _{0}$ in the range $[0.8,1.2]$ and discretize it into $100$ equal elements. We run the algorithm for a number times in order to obtain the excitation probability of the first ancilla qubit. The excitation probability of the ancilla qubit at different eigenvalues of the reference state is shown in Fig.~$1$. From the figure we can see that the excitation probability reaches its maximum at $\varepsilon _{0} =1$, thus the ground state energy of the AKLT model can be obtained as $E_1=0$.

By setting $\varepsilon _{0} =1$, $c=0.05$ and run the algorithm, in the first iteration, when the measurement on the first ancilla qubit is in its excited state, the state we obtained on the last $n$ qubits of the circuit is $|\varphi ^{(1)}\rangle = 0.321 (|0011\rangle + |0101\rangle + |1010\rangle)
- 0.573 (|0110\rangle + |1001\rangle) + 0.186 |1100\rangle $, which has
fidelity of $0.99$ with the ground state of the system. And after running
the algorithm for two iterations, the state we obtained is $|\varphi
^{(2)}\rangle = 0.288 (|0011\rangle + |0101\rangle + |1010\rangle) - 0.577
(|0110\rangle + |1001\rangle) + 0.292 |1100\rangle $, which has fidelity about one with the ground state of the AKLT model.

Here we only use the simplest $1d$-AKLT model as an example to show the
application of the algorithm. In Refs.~\cite{wei, cai}, it was shown that
the ground state of the two-dimensional AKLT model of spin-$3/2$ particles
is a universal resource for one-way quantum computation. In this
model, the spin states of a spin-$3/2$ particle are represented by three
qubits as shown in~\cite{wei}. The $2d$-AKLT model of spin-$3/2$
particles is a gapped model and the algorithm we proposed here can
be applied for preparing its ground state. We can use this algorithm to
obtain its ground state energy, then run the algorithm to
evolve the system to converge to its ground state.

\noindent \textbf{Discussion} \newline
We have proposed a quantum algorithm for preparing the ground state of a system,
provided that the ground state energy is known. A resonance mechanism is introduced
to amplify the amplitude of the ground state that is contained in an initial input
state of the system. A measurement performed on the ancilla qubit indicates
whether the system is prepared to its ground state or not. The procedures of the
algorithm can be run iteratively to drive the system to converge monotonically
to its ground state. This algorithm can be applied to any system with a Hamiltonian
that can be simulated on a quantum computer.

The efficiency of the algorithm depends on the overlap between the initial input state and the ground state of the system, as shown in Eq.~($9$) when the energy gap between the ground and excited states of the system is much larger the coupling coefficient $c$. It is limited by the energy gap between the ground state and excited states, as shown in Eqs.~($6-8$), when the energy gap is small, the amplification rate of the amplitude of the ground state in the input state of the algorithm is limited by the energy gap. If the overlap between the initial input state and the ground state of the system is finite, and the energy gap between the ground state and the first excited state of the system is polynomial large, such that the coupling coefficient $c$ in the algorithm can be set to be much less than the energy gap, then the system can be evolved to its ground state in polynomial time with finite success probability.

We now compare this algorithm with other cooling algorithms for preparing the ground state of a system. Firstly, in our algorithm, the ground state energy of the system is required. For some systems, the ground state energy is already known, such as the example we used and some oracle-based problems in quantum computation. For a system whose ground state energy is unknown, an approximate ground state energy can be obtained using the algorithm in Ref.~\cite{wan} or the algorithm in this work as shown in the example. Obtaining the ground state energy of a system requires some extra cost, but it provides some advantages for the algorithm. By using this algorithm, whether a system is prepared to its ground state can be determined according to the measurement result on the first ancilla qubit. We can be certain that the system is evolved to a state that is close to the ground state of the system when an excitation is observed on the first ancilla qubit. While for other algorithms, there is no direct way of verifying that the system is cooled to its ground state.

Secondly, the algorithm we proposed here can be run iteratively. With the known ground state energy, the system can be evolved to converge monotonically to its ground state through the iterative procedures in the algorithm. This algorithm provides a systematic way of purifying a state to the ground state of a system. While other cooling algorithms do not have such a property.

Thirdly, in our algorithm, a total number of $(n+2)$ qubits are used, while other algorithms use $(n+1)$ qubits. By adding the second ancillary qubit, we can introduce a reference state that does not depend on the energy spectrum of the system, which gives us the flexibility of setting proper resonance transitions to evolve the system to any of its eigenstates. The algorithms in our previous work~\cite{dynamics, wan} that use $(n+1)$ qubits, can also be used for preparing the ground state of a system. But they can only be run in one time and the resonance transition induced depends on the energy spectrum of the system. While in the present algorithm, we can purify the ground state iteratively, the purity of the ground state can be increased in a systematic way.

In Ref.~\cite{taylor}, a cooling algorithm was proposed also based on resonance transition, where a qubit that represents the bath is coupled to the system. The frequency of the qubit is adjusted to match a transition frequency of the system to induce a resonance transition to cool the system. While in our algorithm, by adding the second ancillary qubit, we can introduce a reference state that is independent of the energy spectrum of the system. Its eigenvalue can be adjusted such that the transition frequency between the reference state and the ground state of the system matches the frequency of the first ancilla qubit, therefore a resonance mechanism can be induced. This provides the flexibility of obtaining any eigenstate of the system. Besides, our algorithm can be run iteratively to evolve the system to converge monotonically to its ground state. While the algorithm in Ref.~\cite{taylor} does not have this property.

Adiabatic evolution can also be used for preparing the ground state of a system. In the adiabatic state preparation approach, one starts with an initial Hamiltonian and evolve it adiabatically to the target Hamiltonian, therefore the system is evolved from the ground state of the initial Hamiltonian to the ground state of the target Hamiltonian. While our algorithm starts with the target Hamiltonian directly to obtain its ground state. It requires only information concerning the spectrum of the Hamiltonian of the system, and not any intermediate Hamiltonians of the adiabatic evolution. The implementation is simpler for our algorithm, and whether the system is evolved to its ground state can be determined from the measurement on the first ancilla qubit.

In Ref.~\cite{Hen}, the author introduced a controlled quantum adiabatic evolutions of single- and two-qubit operations for performing universal quantum computation. Compare with the algorithm we proposed here, both algorithms have a similarity in the controlled Hamiltonian, one can see that the Hamiltonian $H_R$ in our algorithm can be thought of as a controlled Hamiltonian evolution. And the initial states for both algorithms are eigenstates of the controlled Hamiltonians, respectively. The difference is that our algorithm is based on a resonance mechanism to obtain the ground state of a system, it starts from the target Hamiltonian directly; while the algorithm in~\cite{Hen}, the desired quantum state is reached through a controlled adiabatic evolution.

\noindent\textbf{Acknowledgements}\newline
This work was supported by the National Nature Science Foundation of
China~(Grants No.~11275145).

\noindent\textbf{Additional information}\newline
Competing financial interests: The authors declare no competing financial
interests.

\noindent Correspondence and requests for materials should be addressed to
Hefeng Wang~(wanghf@mail.xjtu.edu.cn).

\clearpage
\begin{figure}[h]
\begin{center}
\end{center}
\caption{\textbf{Transition frequency spectrum between the reference state, $|\varphi_{0}\rangle$, and the ground state of the AKLT model. The blue solid curve represents the excitation probability of the first ancilla qubit at different eigenvalues of $\varepsilon_0$ of the reference state. The coupling coefficient in Eq.~($1$) $c=0.05$ and the evolution time $\tau =31.4$.}}
\end{figure}

\clearpage

\begin{figure}[h]
\begin{center}
\includegraphics*[width=6in]{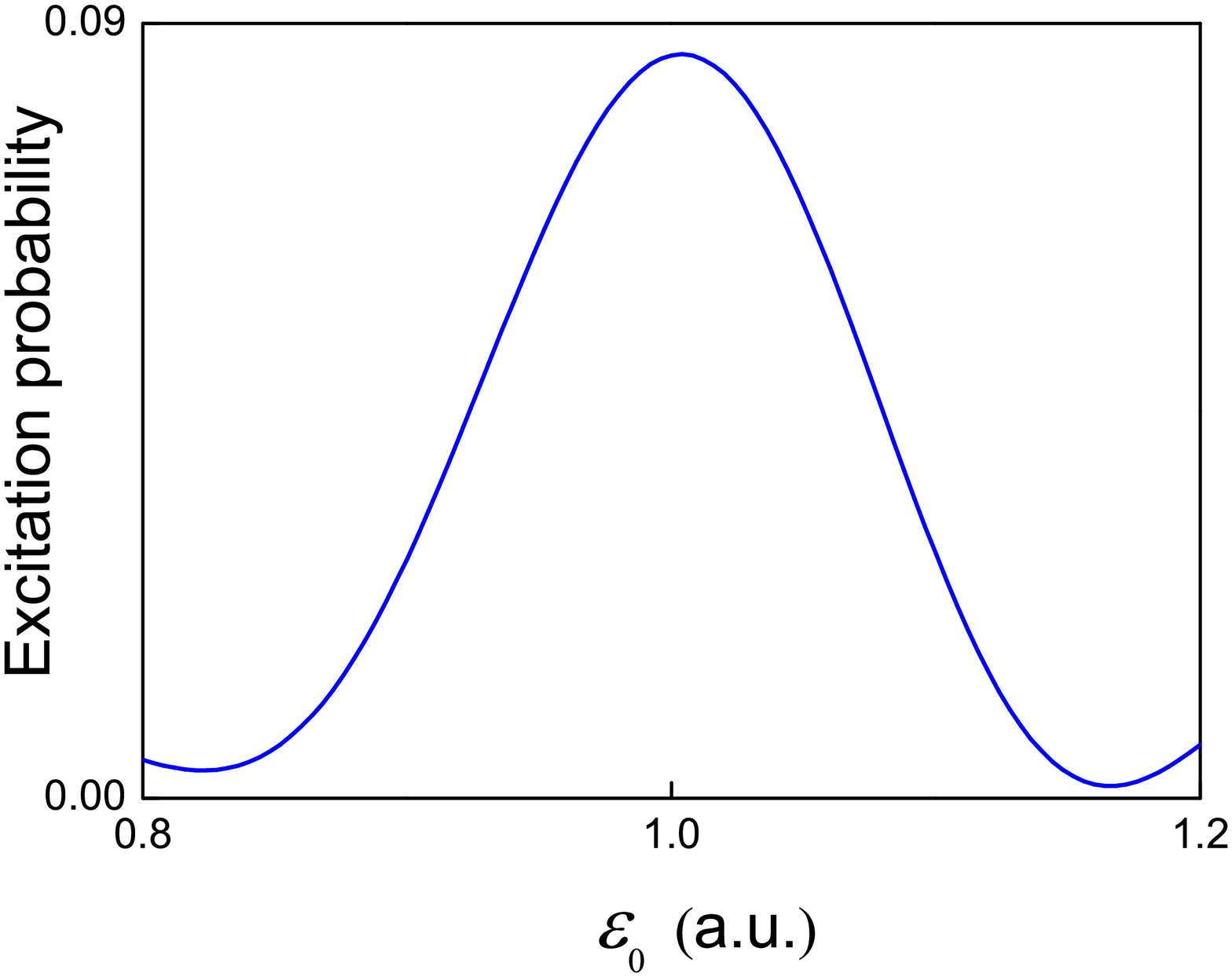}
\end{center}
\end{figure}

\end{document}